\documentclass[12pt]{iopart}
\usepackage{amsfonts}
\usepackage{iopams}
\usepackage{amssymb}
\usepackage{graphicx}
\usepackage{bm}
\usepackage{color}
\usepackage{xcolor}

\newcommand{\beq}{\begin{equation}}
\newcommand{\eeq}{\end{equation}}
\newcommand{\beqa}{\begin{eqnarray}}
\newcommand{\eeqa}{\end{eqnarray}}

\newcommand{\p}{{\cal P}}

\begin{document}

\title[Stable two-dimensional soliton complexes with helicoidal spin-orbit
coupling]{Stable two-dimensional soliton complexes in Bose-Einstein
condensates with helicoidal spin-orbit coupling}
\author{Y. V. Kartashov,$^{1}$ E. Ya. Sherman,$^{2,3}$ B. A. Malomed,$^{4,5}$
and V. V. Konotop$^{6}$}
\address{$^{1}$Institute of Spectroscopy, Russian Academy of Sciences,
Troitsk, Moscow, 108840, Russia}
\address{$^{2}$Department of Physical
Chemistry, The University of the Basque Country UPV/EHU, 48080 Bilbao, Spain}
\address{$^{3}$IKERBASQUE Basque Foundation for Science, Bilbao, Spain}
\address{$^{4}$Department of Physical Electronics, School of Electrical Engineering, Faculty of Engineering,
 and Centre for Light-Matter Interaction, Tel Aviv University, 69978 Tel Aviv, Israel}
\address{$^{5}$Instituto de Alta Investigaci\'{o}n, Universidad de
Tarapac\'{a}, Casilla 7D, Arica, Chile}
\address{$^{6}$Departamento de F\'{i}sica, Faculdade de Ci\^{e}ncias, and Centro de F\'{i}sica Te\'{o}rica e Computacional,
 Universidade de Lisboa, Campo Grande, Edif\'{i}cio C8, Lisboa 1749-016, Portugal}

\eads{\mailto{evgeny.sherman@ehu.es}}

\begin{abstract}
We show that attractive two-dimensional spinor Bose-Einstein condensates
with helicoidal spatially periodic spin-orbit coupling (SOC) support a rich
variety of stable fundamental solitons and bound soliton complexes. Such
states exist with chemical potentials belonging to the semi-infinite gap in
the band spectrum created by the periodically modulated SOC. All these
states exist above a certain threshold value of the norm. The chemical
potential of fundamental solitons attains the bottom of the lowest band,
whose locus is a ring in the space of Bloch momenta,
and the radius of the ring is a non-monotonous function of the SOC strength. The chemical
potential of soliton complexes does not attain the band edge. The complexes
are bound states of several out-of-phase fundamental solitons whose centers
are placed at local maxima of the SOC-modulation phase. In
this sense, the impact of the helicoidal SOC landscape on the solitons is
similar to that of a periodic two-dimensional potential. In particular, it
can compensate repulsive forces between out-of-phase solitons, making their
bound states stable. Extended stability domains are found for complexes
built of two and four solitons (dipoles and quadrupoles, respectively). They
are typically stable below a critical value of the chemical potential.
\end{abstract}

\date{\today}
\maketitle

\noindent\textit{Keywords}: Solitons; matter waves, spin-orbit coupling,
quantum gases

\section{Introduction}

Spin-orbit coupling (SOC) is a fundamentally important effect in physics of
semiconductors, whose theoretical models \cite{Dressel,Rashba} and
experimental manifestations \cite{SOC1,SOC2} have been known for a long
time. More recently, much interest was drawn to emulation of SOC in binary
ultracold gases [chiefly, these are two-component Bose-Einstein condensates
(BEC)] \cite{SOC-BEC1,SOC-BEC2,SOC-BEC3,Zhang_2016}. The SOC emulation
proceeds by mapping the spinor wave function of electrons in semiconductors
into a pseudo-spinor mean-field wave function in BEC, whose components
represent two atomic states in the condensate. Although, in this way,
fermionic spinors are emulated by a pair of hyperfine states of bosonic
atoms, the basic SOC Hamiltonians of the Dresselhaus \cite{Dressel} and
Rashba \cite{Rashba} types, equivalent to non-Abelian or Abelian (depending
on the realization) spin-related gauge fields, can be efficiently reproduced
in BECs illuminated by properly designed sets of laser beams.

While SOC in bosonic gases is a linear effect, its interplay with the
intrinsic BEC nonlinearity, induced by interatomic interactions, was
predicted to give rise to various species of self-trapped mean-field states
\cite{EPL}, including several types of one-dimensional (1D) solitons \cite%
{1Dsol,1Dsol2,1Dsol3,1Dsol4,1Dsol5,KartKon2017,Kartashov_2019_2}. Experimental realization of SOC in an
effectively two-dimensional (2D) geometry was reported too \cite{SOC2D},
which suggests to consider, in particular, a possibility of the creation of
2D gap solitons, supported by a combination of SOC and a spatially periodic
field induced by an optical lattice (OL) \cite{gap-sol}.

A fundamental problem which impedes the creation of 2D and 3D solitons in
BEC, nonlinear optics, and other nonlinear settings, is that the ubiquitous
cubic self-attraction, which usually gives rise to solitons, simultaneously
drives the critical and supercritical collapse (catastrophic
self-compression) in the 2D and 3D cases, respectively \cite{Berge,Fibich}.
Although SOC modifies the conditions of the existence of solutions and of
the blow up, it does not arrest the collapse completely~\cite{Dias_2015}.
The collapse destabilizes formally existing solitons, which makes
stabilization of 2D and 3D solitons a well-known challenging problem \cite%
{SpecialTopics,Kartashov2019_1}. In particular, it is well known that
OL-induced potentials provide a universal method for the creation of stable
multidimensional solitons, including those with embedded vorticity
(topological charge) \cite{Baizakov_2004,Ziad,supervort}. In the absence of
OL potentials, the cubic self-attraction makes vortex solitons extremely
unstable modes \cite{Gaeta,Malomed_2019_1}.

An essential fact, first revealed in Ref. \cite{Sakaguchi_2014}, is that the
standard SOC of the Rashba type, added to the usual two-component system of
Gross-Pitaevskii equations (GPEs) with the attractive cubic interactions [of
both self- and cross-phase-modulation (SPM and XPM) types, i.e., self- and
cross-interactions, respectively], readily supports stable solitons of two
types: semivortices, SVs (with one zero-vorticity component and the other
one with vorticity $1$), and mixed modes, MMs (with vorticities $-1,0$, and $%
1$ blended in both components). Formation of stable states involving nonzero
vorticity can be seen as a result of a counterflow produced by anomalous
spin-dependent velocity determined by SOC \cite{Mardonov_2015}. These SVs
and MMs are stable, severally, in the system with XPM stronger than SPM, and
vice versa. Both SV and MM species are stable in the case of equal strengths
of the SPM and XPM terms (the Manakov's nonlinearity \cite{Manakov_1973}).
When the SV or MM is stable, it represents the system's ground state (in the
absence of SOC, the ground state is missing in the unstable BEC), provided
that the total norm of the pseudo-spinor wave function does not exceed a
critical value (in fact, it is the norm of the \textit{Townes soliton} \cite%
{Chiao_1964}, which is an unstable solution of GPE in 2D in the absence of
SOC \cite{Berge,Fibich}). Above the critical norm, where the centrifugal
counterflow, induced by the anomalous velocity \cite{Mardonov_2015}, cannot
overcome the attraction-induced density flux, the collapse still takes
place, albeit being modified by the SOC. In~\cite{gap-sol} it was shown that
SOC stabilizes solitons, gap-solitons and half-vortices of different
symmetries in Zeeman lattices. Later, similar results for stable 2D solitons
of the SV and MM types were reported for the SOC of a more general type,
combining the Rashba and Dresselhaus couplings \cite{Sakaguchi_2016}, as
well as for the binary BEC of "heavy atoms", for which the kinetic energy
(second derivatives in GPE) may be neglected in comparison with the SOC
energy, represented by first-order cross-derivatives in the GPE system \cite%
{Sakaguchi_2018}. Also demonstrated \cite{Li_2017} were the stabilizations of
MM states by SOC in a system including additional nonlinearity which
represents the beyond-mean-field corrections of the Lee-Huang-Yang type \cite%
{Lee_1957,Petrov_2015_1,Petrov_2016_1,Ferrier_2016,Chomaz_2016,Cabrera_2018,Semengini_2018}. 
Eventually, the creation of
metastable solitons of both SV and MM types was predicted in the 3D system
with SOC \cite{Zhang_2015}. Furthermore, it was demonstrated that the SOC
with reduced dimensionality, \textit{viz}., one- or two-dimensional Rashba
coupling acting in the 2D or 3D space, is sufficient for the stabilization
of the 2D and 3D solitons, respectively \cite{PRR}. The SOC acting in a
spatially confined area is also sufficient for maintaining stable 1D \cite%
{1Dsol5} and 2D \cite{Sandy} solitons.

More generally, the stabilizing effect of gauge fields is known even since
earlier times. In~\cite{Chick} it was shown that stable solitons in
one-component 2D BECs can be sustained by the rotating trap emulating a
gauge field. Confinement of a spinor BEC in the quasi-relativistic limit by
a gauge field was reported in~\cite{Merkl}. Recently, in \cite{KartKon2020}
it was discovered that field components, taken in a pure or non-pure gauge
form, affect the soliton dynamics in very different ways. Localization
characteristics of nonlinear states are determined by the curvature
originating from the non-pure gauge field, while the pure gauge field
strongly affects the stability of the modes. The respective solutions
represent envelopes formed by a non-pure gauge field, which modulates
stationary carrier states created by a pure gauge field. In coupled 1D
nonlinear Schr\"{o}dinger equations with SU(2)-invariant nonlinearity, a
non-uniform pure gauge field preserves the integrability of the system~\cite%
{Kartashov_2019_2}. The total effect of a SOC gauge field can lead not only
to the enhanced stability of solitons in media with attractive interactions,
but also to the existence of novel stable localized modes -- quasisolitons
-- in purely repulsive condensates, both in the free space and in the
presence of traps \cite{KartKon2020}.

In addition to the BEC realm, SOC may be emulated in terms of nonlinear
optics. In that connection, stabilization of 2D spatiotemporal optical
solitons in a dual-core planar waveguide with the help of the counterpart of
the BEC effect, which is represented by temporal dispersion of the
inter-core coupling, was predicted in Ref. \cite{Kartashov_2015}.

Although SOC appears to be a universal mechanism for the stabilization of 2D
solitons, its efficiency is limited. It cannot stabilize excited states,
obtained by adding equal vorticities to both components of SV and MM
solitons in attractive media~\cite{Sakaguchi_2014}, although stable
quasisolitons with embedded vorticity in both components in media with
repulsive cubic interactions have been found~\cite{KartKon2020}.
Possibilities of maintaining bound states of separated solitons by SOC are
not yet known either. For this reason, a relevant possibility, which we
pursue in this work, is to consider SOC in the 2D geometry, subject to
spatially periodic modulation (solitons and their interactions in a 1D BEC
with a helicoidal SOC were considered in \cite{KartKon2017}). In the
experiment, this setting can be implemented by means of an appropriate OL,
which determines the spatially periodic SOC structure. As we demonstrate
below, this system, with a helicoidal orientation of the local SOC [see equation (%
\ref{GrindEQ__1_})], gives rise to a vast stability region for fundamental
2D pseudo-spinor solitons. Further, it also produces stable dipole and
quadrupole bound states of fundamental solitons with opposite signs, which
cannot be found in the absence of the spatially-periodic modulation. We also
demonstrate that stable solitons form sets of states, generated from a seed
one by symmetry transformations.

The rest of the paper is organized as follows. The pseudo-spin system of
coupled GPEs is formulated in Section 2, where we also produce its linear
spectrum, in the numerical form in the general case, and analytically in the
long-wave limit. Symmetries of the system are considered in Section 3.
Systematic numerical results for fundamental solitons, dipoles, and
quadrupoles, including the analysis of their stability and symmetry, are
presented in Section 4. The paper is concluded by Section 5.

\section{The model and linear spectrum of the system}

We describe the evolution of the 2D spinor BEC in the presence of %helicoidal
SOC by coupled GPEs for a spinor wave function ${\bm\Psi }=(\Psi _{1},\Psi
_{2})^{\mathrm{T}}$:
\begin{equation}
i\partial _{t}{\bm\Psi }=\frac{1}{2}[(-i\sigma _{0}\partial _{x}+\alpha ({\bm%
	\sigma }{\bm n}))^{2}+(-i\sigma _{0}\partial _{y}+\alpha ({\bm\sigma }{\bm m}%
))^{2}]{\bm\Psi }-({\bm\Psi }^{\dag }{\bm\Psi }){\bm\Psi }.
\label{GrindEQ__1_}
\end{equation}%
Here ${\bm\sigma}=(\sigma_{1},\sigma_{2},\sigma_{3})$ is the vector of the
Pauli matrices, $\sigma _{0}$ is the identity matrix, $\alpha$ is the
strength of SOC and the position-dependent unit
vectors ${\bm n}$ and ${\bm m}$ determine the SOC direction. We consider a helicoidal SOC with
${\bm n}=(\cos \theta(x,y),0,\sin \theta(x,y))$ and ${\bm m}=(-\sin \theta(x,y),0,\cos \theta(x,y))$.
The above choice of vectors ${\bm m}$ and ${\bm n}$ yields the components 
of a non-Abelian spin-dependent gauge field 
\begin{equation}
 A_{x} =-\alpha \left( \sigma _{1}\cos \theta+\sigma _{3}\sin \theta\right), 
\quad A_{y} =\alpha\left(\sigma _{1}\sin \theta-\sigma_{3}\cos\theta\right)
\end{equation}%
with a position-independent commutator, $[A_{x},A_{y}]=-2i\alpha^{2}\sigma_{2}.$

\begin{figure}[h]
\includegraphics[width=0.99\columnwidth]{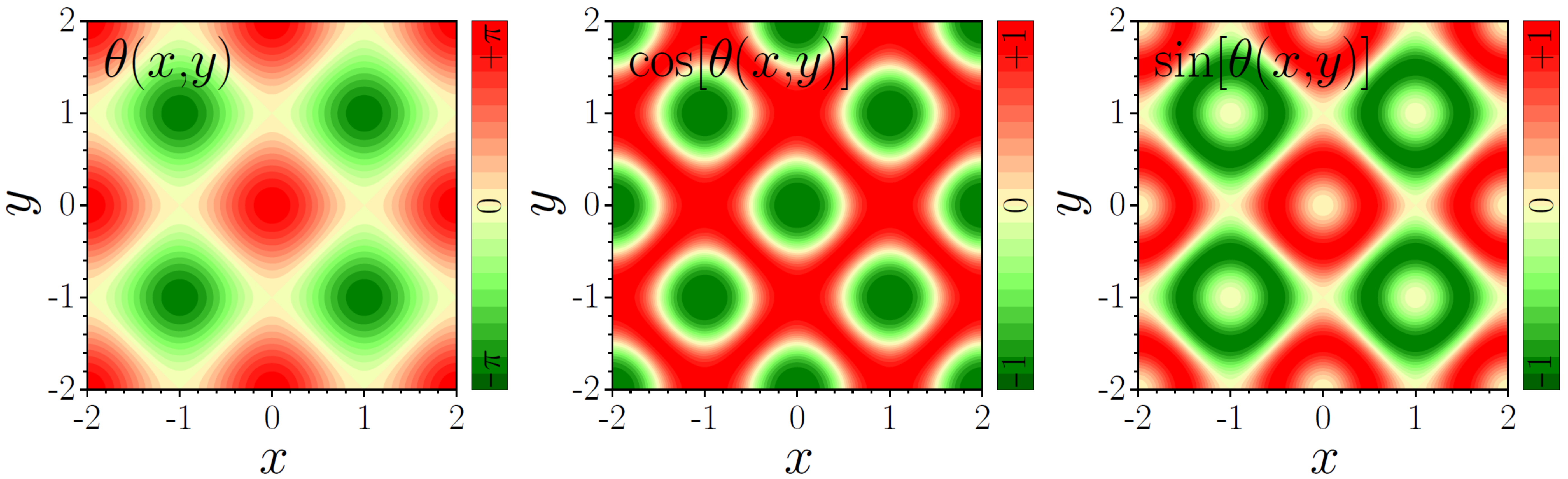}
\caption{ The left panel: contour plot of $\theta(x,y)$, drawn as
per equation (\ref{theta}), with $\lambda _{x}=\lambda %
_{y}=\pi $. Middle and right panels: plots of $\cos\theta(x,y)$ and $\sin\theta(x,y),$ which determine directions of
local vectors ${\bm n}$ and ${\bm m}.$}
\label{figure1}
\end{figure}

In what follows, we concentrate on a spin-orbit coupling realized with a lattice symmetry.  
In this choice the coordinate-dependent 
phase  $\theta(x,y)$ is given by 
\begin{equation}
\theta(x,y)=\frac{\pi }{2}[\cos (\lambda _{x}x)+\cos (\lambda _{y}y)],
\label{theta}
\end{equation}%
with periods $2\pi /\lambda_{x,y}$, as shown in figure \ref{figure1}.

The system includes the attractive nonlinearity with equal strengths of the
self- and cross-interactions in equation~(\ref{GrindEQ__1_}). We stress that no
Zeeman splitting, which may stabilize solitons in other SOC models \cite%
{KartKon2017,KartKon2020}, or external potential is present in equation (\ref%
{GrindEQ__1_}). Although the stabilization of fundamental solitons in the
present setting may be provided by the uniform SOC \cite{Sakaguchi_2014}, it
is demonstrated below that the spatially periodic modulation of local SOC is
necessary to create stable multisoliton complexes (dipoles and quadrupoles),
which do not exist in the uniform space.

The model (\ref{GrindEQ__1_}), (\ref{theta}) has three length
parameters.  Two of them, $2\pi/\lambda_{x,y},$ are the lattice constants. The third one,
$1/\alpha,$ usually referred to as the ``spin-flip length'' and characterizing the coordinate 
dependence of a spinor describing spin 1/2 particle \cite{SOC1} can, in general, be scaled out, i.e., 
the dynamics is governed by $\lambda_{x,y}/\alpha$ ratios. 
Nevertheless, for the analysis of different limiting cases it is convenient to keep $\alpha$ 
and $\lambda_{x,y}$ as separate parameters. 

Before analyzing nonlinear states supported by equation (\ref{GrindEQ__1_}), it
is instructive to consider the spectrum of linear modes determined by
Hamiltonian $H,$ which represents the kinetic energy $H_{\rm kin}$
and spin-orbit coupling $H_{\rm soc}$ in equation (\ref{GrindEQ__1_}) in the form:
\begin{equation}
H=H_{\rm kin} + H_{\rm soc} + \alpha^{2}\sigma_{0}.
\label{Hdefined}
\end{equation}
Here 
\begin{equation}
H_{\rm kin}=-\frac{1}{2}\nabla^{2}\sigma_{0} 
\label{Hkdefined}
\end{equation}
has the spin-diagonal form. The other contribution, $H_{\rm soc},$ can be presented as the sum of two terms:
\begin{equation}
H_{\rm soc}=H_{{\rm soc}|0} + H_{{\rm soc}|1},
\label{Hsocdefined}
\end{equation}
defined as:
\begin{equation}
\hspace{-2cm}H_{{\rm soc}|0} =-i\alpha \left(
\begin{array}{cc}
\sin \theta  & \cos \theta  \\
\cos \theta  & -\sin \theta
\end{array}%
\right)\partial _{x}-
i\alpha \left(
\begin{array}{cc}
\cos \theta  & -\sin \theta  \\
-\sin \theta  & -\cos \theta
\end{array}%
\right) \partial _{y},
\label{Hsoc0defined}
\end{equation}
and 
\begin{equation}
\hspace{-2cm} H_{{\rm soc}|1} = -\frac{i}{2}\alpha (\partial_{x}\theta)
\left(
\begin{array}{cc}
\cos \theta  & -\sin \theta  \\
-\sin \theta  & -\cos \theta
\end{array}%
\right) -\frac{i}{2}\alpha (\partial_{y}\theta)
\left(
\begin{array}{cc}
-\sin \theta  & -\cos \theta  \\
-\cos \theta  & \sin \theta
\end{array}%
\right).  
\label{Hsoc1defined}
\end{equation}%
Hereafter we omitted the explicit spatial dependence  where it is not required directly.

As coefficients in equation (\ref{GrindEQ__1_}) are periodic functions of $x$ and
$y,$ we seek for eigenfunctions in the form of Bloch waves
\begin{equation}
{\bm\Psi }=\left( 
\begin{array}{c}
C_{\uparrow }(x,y)  \\
C_{\downarrow }(x,y)
\end{array}%
\right)
\exp\left[i\left({\bm k}{\bm r} -\mu _{\mathrm{lin}}({\bm k})t\right)\right], 
\label{Bloch}
\end{equation}%
where ${\bm r}=(x,y)$,  $C_{\uparrow,\downarrow }(x,y)=C_{\uparrow,\downarrow }(x+2\pi
/\lambda _{x},y)=C_{\uparrow,\downarrow }(x,y+2\pi /\lambda _{y})$ are
components of a time-independent spatially-periodic spinor wave function, $%
k_{x,y}$ are the Bloch momenta in the first Brillouin zone, $\mu _{%
\mathrm{lin}}({\bm k})$ is the chemical potential of the linear system, and we omitted the Bloch 
band index since we are interested only in the lowest bands. 
The substitution of the Bloch ansatz (\ref{Bloch}) in the linearized version of
equation (\ref{GrindEQ__1_}) leads to an eigenvalue problem,%
\begin{equation}
H\Psi =\mu _{\mathrm{lin}}({\bm k})\Psi,  
\label{HPsi}
\end{equation}
which determines the bandgap spectrum of the system, $\mu_{\mathrm{lin}}({\bm k})$. 

We begin with a particle in the free space, where we can set $\lambda_{x}=\lambda_{y}=0$ and $\theta=\pi,$
the coefficients $C_{\uparrow,\downarrow }(x,y)$ are position-independent, and the values of $k_{x,y}$ are not restricted. 
According to equation (\ref{HPsi}) the chemical potential in this case 
becomes $\mu_{\mathrm{lin}}({\bm k})=\alpha^{2}+k^{2}/2\pm\alpha\,k,$ where $k\equiv(k_{x}^{2}+k_{y}^{2})^{1/2}$
and the spinor components, for different values of $\mu_{\mathrm{lin}}({\bm k}),$ satisfy conditions 
\begin{equation}
|C_{\uparrow}|^{2}=\frac{k \pm k_{y}}{2k},\qquad  |C_{\downarrow}|^{2}=\frac{k \mp k_{y}}{2k}, 
\label{Clin1}
\end{equation}
where upper/lower sign corresponds to the sign in front of $\alpha k$ in $\mu_{\mathrm{lin}}({\bm k}).$

\begin{figure}[h]
\includegraphics[width=0.9\columnwidth]{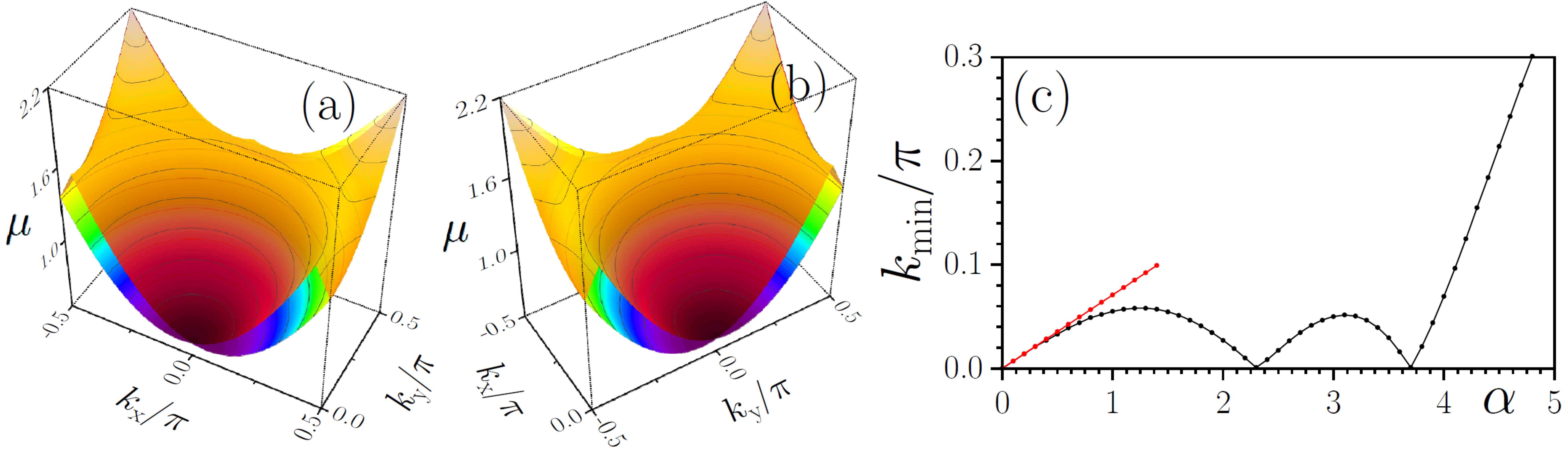}
\caption{(a,b) Two different cuts of $\mu _{\mathrm{lin}}({\bm k})$
dependence in the first Brillouin zone at $\alpha =1$ (two lowest
bands are shown), showing that the minimum of $\mu _{\mathrm{lin}}({
\bm k})$ is achieved on the ring of radius $k_{\min }$. (c) The radius of
the ring versus $\alpha $. The black curve is the numerical result,
while the short red line shows the analytical prediction for $\alpha\ll\,\lambda,$ as given by equation (\ref{kmin}). Note that, at large $%
\alpha,$ the radius $k_{\min}$ increases as $\alpha,$ as expected for the uniform SOC.}
\label{figure2}
\end{figure}

In what follows we concentrate on a square lattice with parameters $\lambda_{x}=\lambda_{y}\equiv \lambda$
and begin with displaying two numerically calculated lowest bands of the linear spectrum in figures %
\ref{figure2}(a,b). The solitons of our interest, formed by the attractive
nonlinearity, belong to the semi-infinite gap with the values of $\mu $ below the
bottom of the lowest band. As one can see from two cuts of the
linear dispersion relation, shown in figures \ref{figure2}(a,b), the minimum
of $\mu_{\mathrm{lin}}({\bm k})$ is achieved on the ring of radius $k_{\min}$ 
in the $(k_{x},k_{y})$ plane. The radius of this ring, being a non-monotonous
function of $\alpha $, shown in figure \ref{figure2}(c) by the black curve, may shrink to 
zero at certain values of $\alpha,$  while it monotonously increases at a sufficiently strong SOC.

For small Bloch momenta, $\left\vert k\right\vert \ll \lambda$,
and small SOC strength, $\alpha\ll\lambda$, one can find the spectrum in an
approximate analytical form by means of averaging with respect to relatively rapid
oscillations of $\cos\theta(x,y)$ and $\sin\theta(x,y)$, while keeping the slowly
oscillating function $\exp\left(i{\bm k}{\bm r}\right)$ in
ansatz (\ref{Bloch}). A nontrivial result can be obtained in the asymptotic
regime defined by the following relation between the two small parameters: 
$k\sim\alpha.$ Under this condition, in the
zero-order approximation the spinor wave function in equation (\ref{Bloch}) does
not depend on the coordinates, $C_{\uparrow,\downarrow}^{(0)}=\mathrm{const}$, 
an estimate for the lowest-order corrections to it being $\delta\left(C_{\uparrow,\downarrow }\right)\sim\alpha k/\lambda^{2}.$ 
Then, the kinetic energy term in Hamiltonian (\ref{Hkdefined}) yields an obvious
lowest-order contribution to the eigenvalue, that is $k^{2}/2$. In
the same approximation, a contribution of terms $\sim i\alpha \partial _{x,y}$ 
in (\ref{Hsocdefined}) can be found as eigenvalues of the respective matrix
with entries, $\cos\theta(x,y)$ and $\sin\theta(x,y)$, replaced by their spatially
average values, $\left\langle \cos \theta(x,y) \right\rangle \equiv \mathrm{{%
\mathcal{C}}}$ and $\left\langle \sin \theta(x,y) \right\rangle \equiv \mathrm{{%
\mathcal{S}}}$, and $-i\partial _{x,y}$ replaced by $k_{x,y}$, the result
being $\pm \alpha \sqrt{\mathrm{{\mathcal{C}}}^{2}+\mathrm{{\mathcal{S}}}^{2}}k$. 
As concerns corrections to the eigenvalues produced by corrections to
the eigenfunctions given by this $\delta\left(C_{\uparrow,\downarrow }\right)$, they scale as 
$\delta\mu\sim\left(\alpha k\right)^{2}$, being negligible in comparison with
the terms written above. In
fact, the current approximation is similar to the lowest-order perturbation
theory in quantum mechanics, which produces the first contribution to the
eigenvalue of energy, induced by a perturbation Hamiltonian, omitting a
correction to the stationary wave function.

Thus, collecting the above terms, the lowest-order approximation for the
spectrum at small $k$ and $\alpha$ is obtained as%
\begin{equation}
\mu _{\mathrm{lin}}({\bm k})-\mu _{\mathrm{lin}}(0)=\frac{k^{2}}{2}\pm \alpha
\sqrt{\mathrm{{\mathcal{C}}}^{2}+\mathrm{{\mathcal{S}}}^{2}}k.
\label{result1}
\end{equation}%
This result is written with respect to $\mu _{\mathrm{lin}}(0)$ because the calculation
of $\mu _{\mathrm{lin}}(0)$ is more cumbersome, as the contribution to it from
the last two terms in Hamiltonian (\ref{Hdefined}) requires a calculation of
the first correction to the spinor eigenfunction, the result being $\mu _{%
\mathrm{lin}}(0)=\alpha ^{2}\left(1+{\mathcal{C}}^{2}\right)/2$. 

For the SOC modulation in equation (\ref{theta}) with $\lambda=\pi$, a
straightforward calculation using the well-known decomposition $\exp (iz\cos\zeta)=J_{0}(z)+\sum_{l=1}^{\infty
}i^{k}J_{l}(z)\cos (l\zeta )$ \cite{Abramowitz} with Bessel functions $J_{l}(z)$ yields $\mathrm{{\mathcal{C}}}=J_{0}^{2}\left(
\pi /2\right)\approx\, 0.223$ and $\mathrm{{\mathcal{S}}}=0.$
Therefore, the analytically predicted spectrum (\ref{result1}) takes the following form:%
\begin{equation}
\mu _{\mathrm{lin}}(k)=\frac{k^{2}}{2}\pm \alpha J_{0}^{2}\left( \frac{\pi }{%
2}\right) k+\frac{\alpha ^{2}}{2}\left[ 1+J_{0}^{4}\left( \frac{\pi }{2}%
\right) \right] . \label{GrindEQ__2_}
\end{equation}%
In the same approximation, the structure of the constant spinor eigenstate
in equation (\ref{Bloch}) is determined by the ratio of its components (cf. equation (\ref{Clin1})):%
\begin{equation}
\frac{C_{\uparrow }^{(0)}}{C_{\downarrow }^{(0)}}=\frac{\mathrm{{\mathcal{C}}%
}k_{y}+\mathrm{{\mathcal{S}}}k_{x}\pm \sqrt{\mathrm{{\mathcal{C}}}^{2}+%
\mathrm{{\mathcal{S}}}^{2}}k}{\mathrm{{\mathcal{C}}}k_{x}-\mathrm{{\mathcal{S%
}}}k_{y}}\equiv \frac{k_{y}\pm k}{k_{x}}.  \label{eigen}
\end{equation}
The branch of dispersion relation (\ref{GrindEQ__2_}) with the bottom sign has a minimum,
$\mu _{\min }=\alpha ^{2}/2$ at a finite Bloch momentum,%
\begin{equation}
k_{\min }=\alpha J_{0}^{2}\left( \pi /2\right) .  \label{kmin}
\end{equation}%
The derivative $dk_{\min }/d\alpha =J_{0}^{2}\left( \pi /2\right)$
exactly coincides with its numerically found counterpart at $\alpha\rightarrow 0$, 
as shown in figure \ref{figure2}(c). The same figure suggests
that the analytical approximation remains accurate enough up to $\alpha
\simeq 0.5$, which is corroborated by a detailed comparison with numerical
data (not shown here).

Finally, the exact numerical calculation of $\mu$ at the
bottom of the Bloch band, demonstrates that it is equal $\alpha^{2}/2$ for all $\alpha.$
Therefore, the relevant for soliton existence band edge $\mu_{\rm be}=\alpha^{2}/2.$

\section{Symmetry properties of solitons}

Now we turn to exact soliton solutions of equation~(\ref{GrindEQ__1_}) in the
form of ${\bm\Psi }=e^{-i\mu t}\bpsi(x,y)$, with $t$-independent complex
spinor $\bpsi=\left( \psi _{1}(x,y),\psi _{2}(x,y)\right) ^{\mathrm{T}}.$
Thus, we consider the nonlinear eigenvalue problem:
\begin{equation}
\mu \bpsi=H\bpsi-\left(\bpsi^{\dag}\bpsi\right) \bpsi,
\label{GrindEQ_stat}
\end{equation}%
where the linear part $H$ of the model's Hamiltonian (\ref{GrindEQ__1_}) is
given by equation (\ref{Hdefined}). The consideration below is limited to the case of a square 
lattice cell with equal $x-$ and $y-$periods: $\lambda_{x}=\lambda _{y}=\lambda$, and hence the general symmetry properties pertain to this choice. 
To describe families of nonlinear solutions
(which in our case do not bifurcate from the linear limit), we first address
the symmetries of $H$. To this end we introduce time reversal operator $%
\mathcal{T}$ acting on $\bpsi$ as the complex conjugation, $\mathcal{T}\bpsi=%
\bpsi^{\ast }$, $x-$ and $y-$ spatial-inversion operators $\mathcal{P}_{x}$
and $\mathcal{P}_{y}$, acting as
\begin{equation}
\mathcal{P}_{x}\bpsi(x,y)=\bpsi(-x,y),\quad \mathcal{P}_{y}\bpsi(x,y)=\bpsi%
(x,-y),
\end{equation}%
as well as $\mathcal{P}=\mathcal{P}_{x}\mathcal{P}_{y}$ mapping, $%
(x,y)\rightarrow (-x,-y)$, the exchange operator,
\begin{equation}
J\bpsi(x,y)=\bpsi(y,x),
\end{equation}
and two operators of translations by half-period along the $x$ and $y$
directions:
\begin{equation}
\tau _{x}\bpsi(x,y)=\bpsi(x+\pi /\lambda ,y),\qquad \tau _{y}\bpsi(x,y)=\bpsi%
(x,y+\pi /\lambda ).
\end{equation}%

Now we can identify the symmetries of $H$. First, $H$ is $\mathcal{PT}$%
-symmetric: $[H,\mathcal{PT}]=0$. The second symmetry, $\beta ,$ commuting
with the linear Hamiltonian, $[H,\beta ]=0$, can be constructed as
\begin{equation}
\beta =\mathcal{P}_{x}J\left( \sigma _{0}+i\sigma _{2}\right) /\sqrt{2}%
,\qquad \sigma _{0}+i\sigma _{2}=\left(
\begin{array}{cc}
1 & 1 \\
-1 & 1%
\end{array}%
\right) .  \label{beta}
\end{equation}%
Using properties $\mathcal{P}_{x}J\mathcal{P}_{x}=\mathcal{P}_{y}J\mathcal{P}%
_{y}=\mathcal{P}J=J\mathcal{P}$, one can verify that $\beta $ is a generator
of a cyclic group of order $8$ ($\beta^{8}=1$), which is isomorphic to $\mathbb{Z}_{8} $ \cite{Knox,Kurzweil}. Both $\mathcal{PT}$ and $\beta $ symmetries do not
contain shifts, hence they are referred to as \emph{local symmetries}. 
Meantime we observe that there exists also the discrete translational symmetry over the primitive
lattice vectors, i.e., ${\bm u}_{1}=(2,0)$ and ${\bm u}_{2}=(0,2)$ in our
case, as well as the symmetries related to the shifts over half-period and defined 
as $\sigma_3\p_x\tau_x\tau_y$ and $\sigma_1\p_y\tau_x\tau_y$. Since here we are 
interested only in localized solutions, these lattice translations will not be considered.

Taking into account that $\beta ^{4}=-1$, one identifies
\begin{equation}
\mathfrak{G}_{\mathrm{loc}}=\{\pm 1,\pm \mathcal{PT},\pm \mathcal{T}_{F},\pm
\beta ,\pm \beta ^{2},\pm \beta ^{3},\pm \beta \mathcal{PT},\pm \beta ^{3}%
\mathcal{PT}\}  \label{Gloc}
\end{equation}%
as the group of local symmetries of $H$. We emphasize that although the group $\mathfrak{G}_{\mathrm{loc}}$ 
is obtained for the particular choice of $\theta(x,y),$ the following analysis 
is not restricted to this choice and depends only on the group $\mathfrak{G}_{\mathrm{loc}},$ 
which can be realized for a broad variety of the $H-$Hamiltonians.
In $\mathfrak{G}_{\mathrm{loc}}$ we use notation $\mathcal{T}_{\mathrm{F}}=\beta ^{2}\mathcal{PT}=i\sigma _{2}%
\mathcal{T}$ for the spin-1/2 time reversal, and refer to elements with "$+$%
" ("$-$") sign as symmetries (anti-symmetries). Then, one can distinguish
three cases as follows. If, for a given solution $\bpsi$ of equation (\ref%
{GrindEQ_stat}), there exists an element $g$ of $\mathfrak{G}_{\mathrm{loc}}$
such that $\bpsi_{g}=g\bpsi$ is a solution of equation (\ref{GrindEQ_stat}) with
the same $\mu $, we define $\bpsi_{g}$ as a \emph{$g$-transformed} solution.
If, additionally, $\bpsi_{g}=\bpsi$ (or $\bpsi=g\bpsi$), such a solution is
termed a \emph{$g$-symmetric} one. If a given solution features $\bpsi\neq g%
\bpsi$ for all $g\in \mathfrak{G}_{\mathrm{loc}}$, then one says that
\textit{symmetry breaking} occurs. Although we cannot exclude a possibility
of the symmetry breaking in our case, the extensive numerical studies of
localized states of equation (\ref{GrindEQ_stat}) have not revealed such a
possibility. Therefore, we focus below on the existence and properties of
"symmetric" solutions, i.e., ones generated by all symmetry transformations
from $\mathfrak{G}_{\mathrm{loc}}$.

Generally speaking, a symmetry of a linear Hamiltonian does not yet imply
the symmetry of the complete nonlinear Hamiltonian. For a nonlinear solution
to be generated by a symmetry $g$, the nonlinearity must be compatible with
that symmetry. There exist two possibilities for that to occur~\cite{ZezKon}%
. Either the nonlinearity is $g$-symmetric, i.e., in our case, $[g,\bpsi%
^{\dag }\bpsi]=0$ for any $\bpsi$, or it is \emph{weakly} $g$-symmetric,
i.e., $\bpsi^{\dag }\bpsi=(g\bpsi)^{\dag }g\bpsi$ is verified for a $g$%
-transformed solution $g\bpsi$ (although not necessarily for any $\bpsi$).
It is straightforward to verify that the Manakov's (SU(2)-symmetric)
nonlinearity in equation (\ref{GrindEQ_stat}) is weakly symmetric with respect to
local elements of $\mathfrak{G}_{\mathrm{loc}}$. Therefore, below we look
for nonlinear modes generated by the symmetry group $\mathfrak{G}_{\mathrm{%
loc}}$.

%Suppose that a spinor $\bpsi$ solves (\ref{GrindEQ_stat}).
Using properties of the cyclic group of the $\beta $ transformations, one
concludes that there exist no nontrivial $\beta ^{n}$-symmetric solutions.
Indeed, starting with $\beta ^{2}$, the existence of $\beta ^{2}$-symmetric
solution $\bpsi_{\beta ^{2}}$ implies that $-\bpsi_{\beta ^{2}}=\beta ^{4}%
\bpsi_{\beta ^{2}}=\beta ^{2}\bpsi_{\beta ^{2}}=\bpsi_{\beta ^{2}}$, i.e., $%
\bpsi_{\beta ^{2}}\equiv 0$. Further, if a solution is $\beta $-symmetric,
it is also $\beta ^{2}$-symmetric, and hence is zero, too. These arguments
are repeated for all $\pm \beta ^{n}$-symmetries. Hence, $\beta ^{n}$
transformation generates new solutions: if any solution $\bpsi$ of (\ref%
{GrindEQ_stat}) is found, then $\beta ^{n}\bpsi\neq \bpsi$ (for $%
n=1,\ldots,7 $) is another solution corresponding to the same $\mu $.
% (notice that $[\PT,\beta]=0$).

Not all symmetries from $\mathfrak{G}_{\mathrm{loc}}$ result in \emph{%
physically distinct} solutions. For any symmetry
%(sign "$+$" in $\mathfrak{G}$)
involving antilinear time reversal $\mathcal{T}$, the anti-symmetry
%(sign "$-$" in $\mathfrak{G}$)
amounts to a trivial phase shift. For instance, if solution $\bpsi_{\mathcal{%
PT}}$ is a $\mathcal{PT}$-symmetric one, then $i\bpsi_{\mathcal{PT}%
}$ is anti-$\mathcal{PT}$-symmetric. Therefore, in the analysis of symmetric
solutions of equation (\ref{GrindEQ_stat}) presented below, we exclude all
anti-symmetries. This leaves us with only \emph{four} transformations: $%
\mathcal{PT}$, $\mathcal{T}_{F}$, $\beta \mathcal{PT}$, and $\beta ^{3}%
\mathcal{PT}$, that may generate physically different solutions. Among these
elements only the $\mathcal{PT}$-symmetry allows for symmetric solutions: $%
\bpsi_{\mathcal{PT}}=\mathcal{PT}\bpsi_{\mathcal{PT}}$. Now, let us consider
$\bpsi_{\mathcal{PT}}$ and its $\beta ^{2}$-transformation: $\bpsi_{\mathcal{%
T}_{F}}=\beta ^{2}\bpsi_{\mathcal{PT}}=\mathcal{T}_{F}\bpsi_{\mathcal{PT}%
}=i\sigma _{2}\mathcal{P}\bpsi_{\mathcal{PT}}$. One readily concludes that
the absolute value of the first (second) component of $\bpsi_{\mathcal{T}%
_{F}}$ is the $\mathcal{P}$-inverse absolute value of the second (first)
component of $\bpsi_{\mathcal{PT}}$. In this sense, these solutions are
topologically similar. The topological similarity is also verified for the
pair of $\beta \mathcal{PT}$- and $\beta ^{3}\mathcal{PT}$-symmetric
solutions related by the $\beta ^{2}$ transformation. On the other hand, $%
\mathcal{PT}$- and $\beta \mathcal{PT}$-symmetric solutions have essentially
different distributions of the fields in both components.

The above analysis is applicable for any family of localized nonlinear
modes. Although the detailed analysis of numerically obtained families
(fundamental, dipole-, and quadrupole-like) will be done in the next
Section, here we can illustrate their main symmetry-related properties.
These solitons, found by the Newton relaxation method, that permits
obtaining entire families parameterized by the chemical potential $\mu ,$
are characterized by their norm $U(\mu )$ and size $w(\mu )$, defined as:
\begin{equation}
U(\mu )\equiv \int \bpsi^{\dag }\bpsi d^{2}r;\qquad w(\mu )\equiv {U(\mu )}%
\left[ \int (\bpsi^{\dag }\bpsi)^{2}d^{2}r\right] ^{-1/2}
\end{equation}%
and the peak amplitudes of spinor components $a_{1,2}(\mu)=\max \left\vert
\psi _{1,2}\right\vert .$

\begin{figure}[h]
\includegraphics[width=0.9\columnwidth]{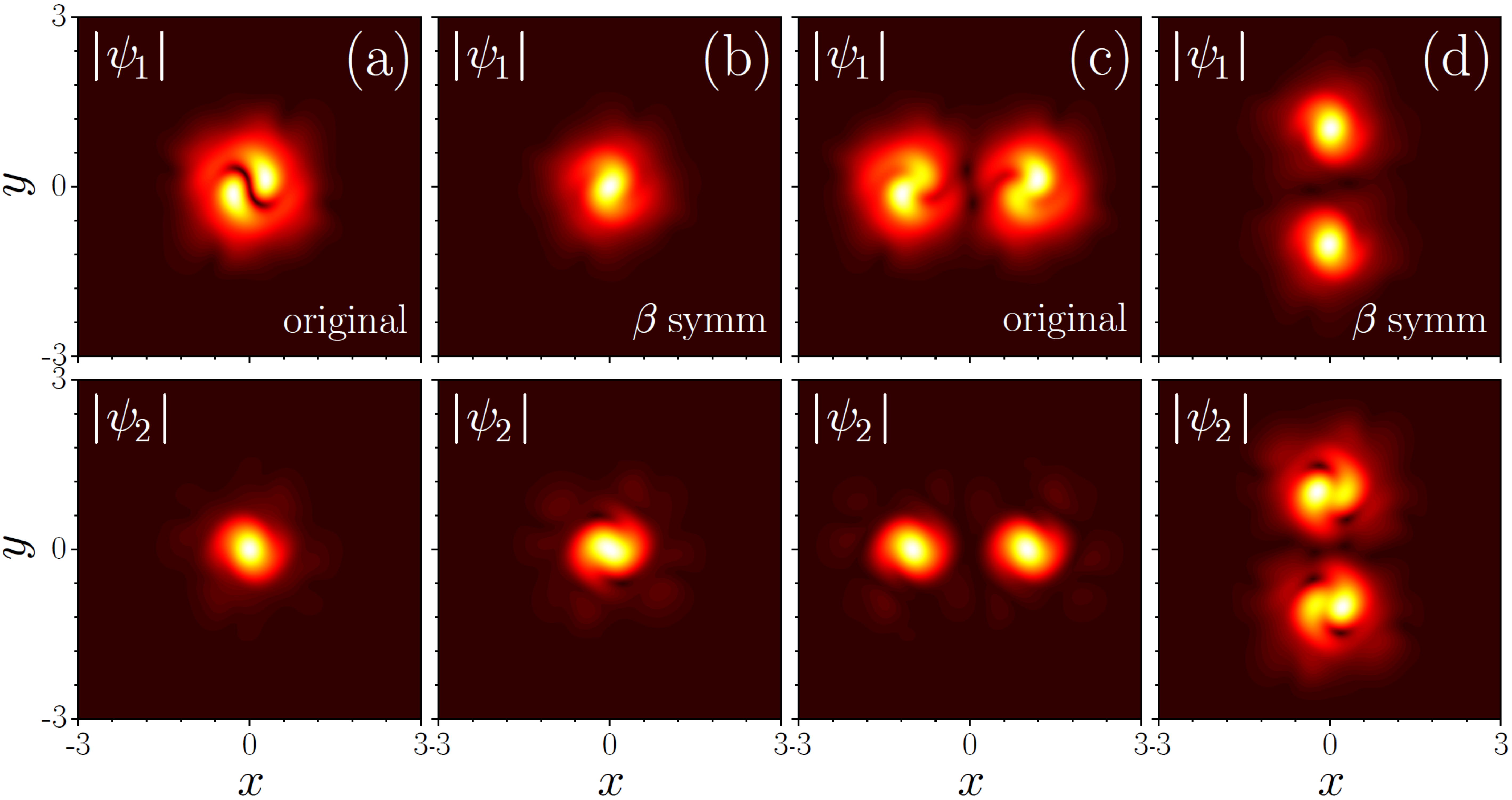}
\caption{Examples of stable $\mathcal{PT}$-symmetric solutions and their $%
\beta $-transformations for the branches of fundamental (a,b) and
dipole (c,d) solitons at $\mu =2$, $\alpha =3$.}
\label{figure3}
\end{figure}

In figure~\ref{figure3} we show $\mathcal{PT}$-symmetric solutions (panels (a)
and (c)) and the respective $\beta $-transformed solutions (panels (b) and
(d)) for the families of fundamental solitons (panels (a) and (b)) and
dipoles (panels (c) and (d)). It is straightforward to verify that both $%
U(\mu )$ and $w(\mu )$ are equal for a given $\mathcal{PT}$-symmetric
soliton and its $\beta ^{n}$-transformation. It is also clear that the
linear stability of a $\mathcal{PT}$-symmetric state coincides with the
stability of its $\beta ^{n}$ transformation: if $\bpsi_{\mathcal{PT}}$ is
stable (unstable), the same is valid for $\beta ^{n}\bpsi_{\mathcal{PT}}$.
Therefore, the nonlinear modes shown in figures~\ref{figure3} (a) and (b) (and
in figures~\ref{figure3} (c) and (d)) are either both linearly stable or both
unstable, in the former case being stabilized by the nonlinearity.

\section{Soliton families and their stability: numerical results}

\begin{figure}[h]
\includegraphics[width=0.9\columnwidth]{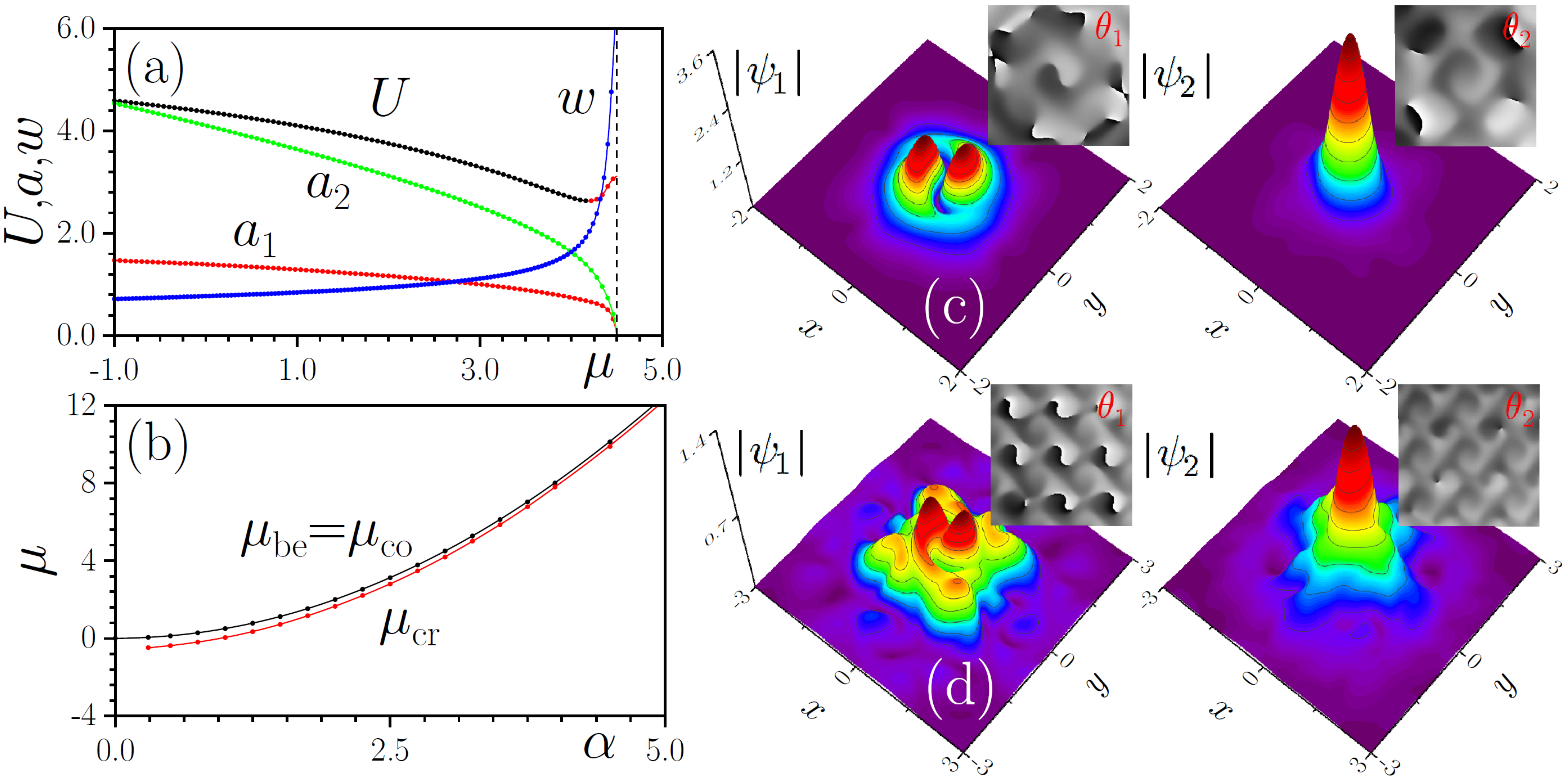}
\caption{(a) Norm $U(\mu )$, size $w(\mu )$, and component
amplitudes, $a_{1,2}(\mu)$, of fundamental solitons versus chemical potential $%
\mu $ at $\alpha =3$. Black and red segments in the $U(%
\mu )$ dependence correspond to stable and unstable states,
respectively. The dashed line shows the location of the bottom of the first
band. (b) Existence and stability domains of the fundamental solitons in the
$(\alpha ,\mu )$ plane. The solitons exist at $\mu %
\leq \mu _{\mathrm{be}}=\mu _{\mathrm{co}}$, and are stable
at $\mu \leq \mu _{\mathrm{cr}}$. Panels (c) and (d) display
examples of profiles of the absolute value and phase (insets) of the
fundamental solitons with $\mu =1$ (c, stable) and $\mu =4.2$
(d, taken at the stability border) are shown for $\alpha =3$.}
\label{figure4}
\end{figure}

The simplest family of self-trapped solutions supported by the 2D helicoidal
SOC landscape is represented by fundamental solitons. To be stable, such
solitons have to be centered at a maximum of the periodic SOC phase, $\theta(x,y)=\pi,$
with the chemical potential taking values below the edge of the
first band in the linear spectrum, i.e., at $\mu \leq \mu _{\mathrm{be}}$
[see the dashed line in figure \ref{figure4}(a)]. The dominant component
(either $\psi _{1}$ or $\psi _{2}$, see Sec. 3) of such a soliton typically
has a bell-shaped profile with weak modulations on the top of it, while a weaker
component may feature a very complex profile with curved nodal lines, see,
e.g., figure \ref{figure4}(c). Both components of the wave function have
nontrivial phase distributions, as shown in the insets of figure \ref{figure4}%
(c). When the chemical potential approaches the bottom of the first linear band,
the soliton expands and its width diverges, see a
representative $w(\mu )$ dependence in figure \ref{figure4}(a). In this
regime, spatial modulations of the soliton shape become much more pronounced for
both components [figure \ref{figure4}(d)]. Peak amplitudes $a_{1,2}(\mu)$ of both
components decrease with the increase of $\mu $ and vanish exactly at the
edge of the allowed band [figure \ref{figure4}(a)]. In contrast, the soliton's
norm, $U(\mu) ,$ shows a non-monotonous behavior, clearly indicating that
solitons exist only above a certain threshold value of the norm. This threshold 
decreases with the increase of the SOC strength $\alpha $.
At $\alpha \rightarrow 0$, the dependence $U(\mu )$ approaches a constant
value, $U_{T}\approx 5.85$, which is the above-mentioned Townes' soliton
norm \cite{Berge,Fibich,Chiao_1964}.

The growth of deformation and non-monotonous character of the $U(\mu )$
dependence appearing with the increase of $\alpha $ suggests that the helicoidal SOC
may play a stabilizing role for fundamental solitons. We have checked the
stability of such states by simulating equation (\ref{GrindEQ__1_}) with small
input complex noise added to stationary profiles, up to very large times. As
a result, it was found that large parts of $U(\mu )$ families, with negative
slope $\partial U(\mu)/\partial \mu $, represent stable fundamental solitons.
These findings are summarized in the existence-and-stability diagram
displayed in figure \ref{figure4}(b), where the solitons exist at $\mu \leq
\mu _{\mathrm{be}}$, and are stable for all values of $\mu \leq \mu _{%
\mathrm{cr}}$, the border of the stability domain exactly corresponding to
the $\left. \partial U(\mu )/\partial \mu \right\vert _{\mu =\mu _{\mathrm{cr%
}}}=0$ boundary, in agreement with the well-known Vakhitov-Kolokolov
criterion \cite{Berge,Fibich,VK}.

The existence of stable fundamental solitons residing at the maxima of the
periodic $\theta(x,y)$ function suggests that equation (\ref{GrindEQ__1_})
allows to construct complexes of the solitons, which cannot be done with the
help of spatially-uniform SOC \cite%
{Sakaguchi_2014,Sakaguchi_2016,Sakaguchi_2018}. We have found that it is
indeed possible to create stable complexes of spatially separated solitons
with opposite signs of ${\bm\psi}$, i.e., dipoles and quadrupoles.

\begin{figure}[h]
\includegraphics[width=0.9\columnwidth]{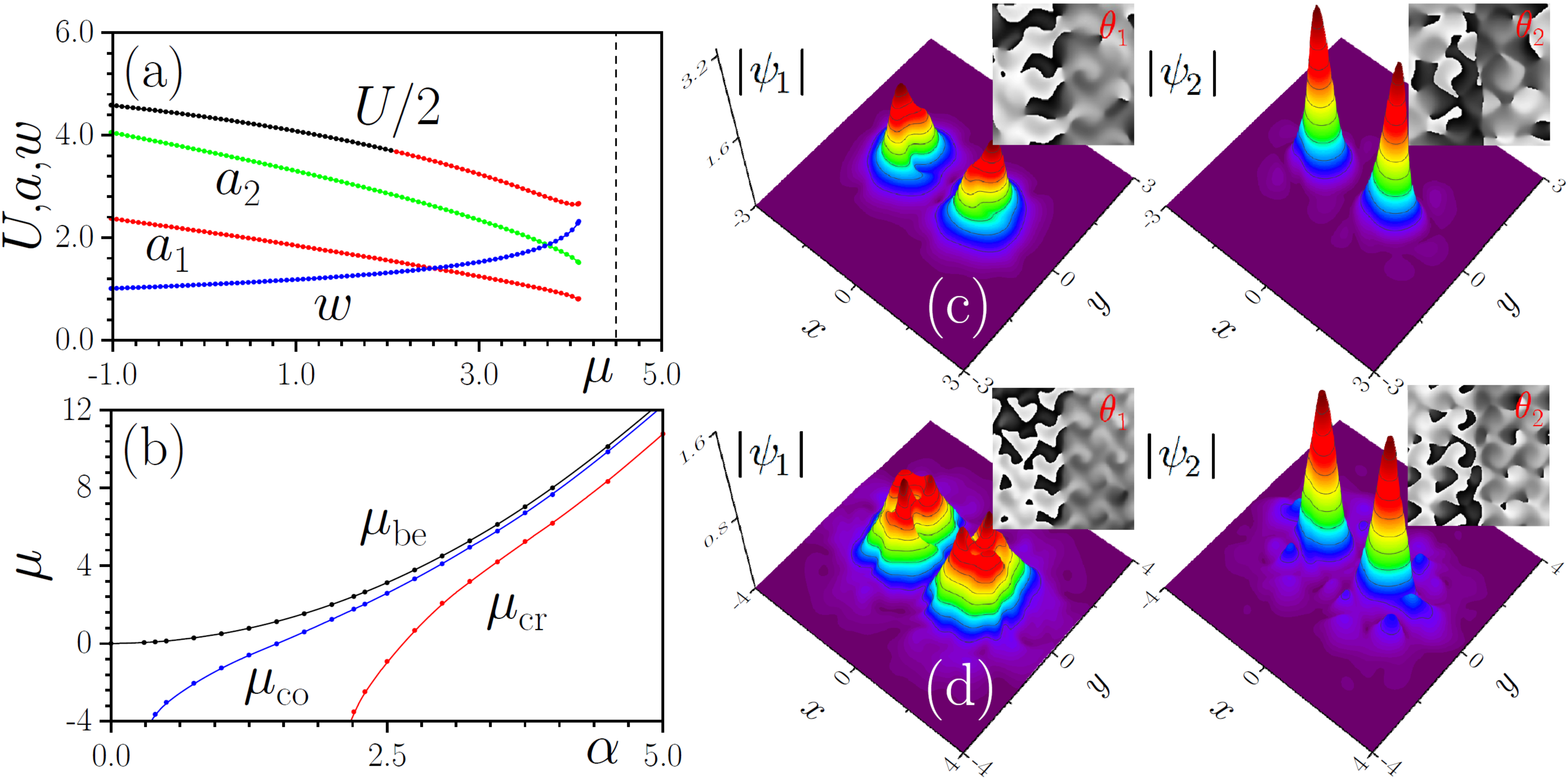}
\caption{(a) The norm $U(\mu )$, size $w(\mu )$, and
component amplitudes, $a_{1,2}(\mu )$, of dipole solitons versus chemical
potential $\mu $ at $\alpha =3$. Black and red segments in
the $U(\mu )$ dependence correspond to stable and unstable states,
respectively. The dashed line shows the location of the bottom level of the
first band. (b) Existence and stability domains of the dipole solitons in
the $(\alpha ,\mu )$ plane. They exist at $\mu \leq
\mu _{\mathrm{co}}<\mu _{\mathrm{be}}$, and are stable at $%
\mu \leq \mu _{\mathrm{cr}}$. Panels (c) and (d) display
examples of profiles of the absolute value and phase (insets) of dipole
solitons with $\mu =1$ (c, stable) and $\mu =4$ (d,
unstable) for $\alpha =3$. For convenience for dipole solitons the
SOC profile was shifted by half of the period $\pi/\lambda$
along the $x$-axis.}
\label{figure5}
\end{figure}

Examples of the simplest dipole soliton complex, that may be considered as a
pair of two solitons with opposite signs, and properties of dipole
families are presented in figure \ref{figure5}. The dipole is built of two
soliton-like density maxima placed at adjacent maxima of $\theta(x,y)$ and,
accordingly, separated by distance $\approx 2\pi /\lambda =2.$ The solitons
interact through their tails, the interaction getting stronger with the
increase of chemical potential $\mu $, due to weakening localization of each
soliton [cf. figures \ref{figure5}(c) and (d)]. Even though phase
distributions in the dipole modes are quite complex, due to spatially
dependent SOC, one can see in the insets that the two density maxima in both
components have, \textquotedblleft on average\textquotedblright , phase
shifts of $\pi $, which corresponds to the opposite signs of the two
solitons. Because dipole solitons are excited nonlinear states, their
chemical potential does not reach the bottom of the band. Instead, the
dipoles exist below a certain cutoff value, $\mu =\mu _{\mathrm{co}}$, whose
$\alpha-$dependence is shown in figure \ref{figure5}(b) by
the blue line. At the cutoff position, amplitudes of both components of the
dipole's wave function, $\psi _{1,2}$, do not vanish, and the width remains
finite, but dependence $U(\mu )$ acquires a vertical tangential direction
[for illustrative purposes, in figure \ref{figure5}(a) we plot half of total
norm, as it is comparable with that of fundamental soliton]. Because the
capability of SOC landscape to compensate repulsive forces acting between
solitons with opposite signs reduces with the decrease of $\alpha $, the
cutoff value $\mu _{\mathrm{co}}$ rapidly decreases at $\alpha \rightarrow 0$
[figure \ref{figure5}(b)].

An important finding is that the dipole solitons enabled by helicoidal SOC
may be stable. The stabilization is possible below a critical value of the
chemical potential, $\mu \leq \mu _{\mathrm{cr}}$ [see the stable (black)
segment of the $U(\mu )$ branch in figure \ref{figure5}(a)], while, close to
the cutoff, the dipoles are unstable [see the red segment of the $U(\mu )$
branch]. The entire existence and stability domains for the dipole solitons
is presented in figure \ref{figure5}(b): they exist at $\mu \leq \mu _{\mathrm{%
co}}$, and are stable at $\mu \leq \mu _{\mathrm{cr}}$. In the instability
domain, the most typical destabilization mechanism is spontaneous breakup of
symmetry between two density maxima, accompanied by oscillations with
a growing amplitude, leading to the collapse or formation of a fundamental
soliton.

\begin{figure}[h]
\includegraphics[width=0.9\columnwidth]{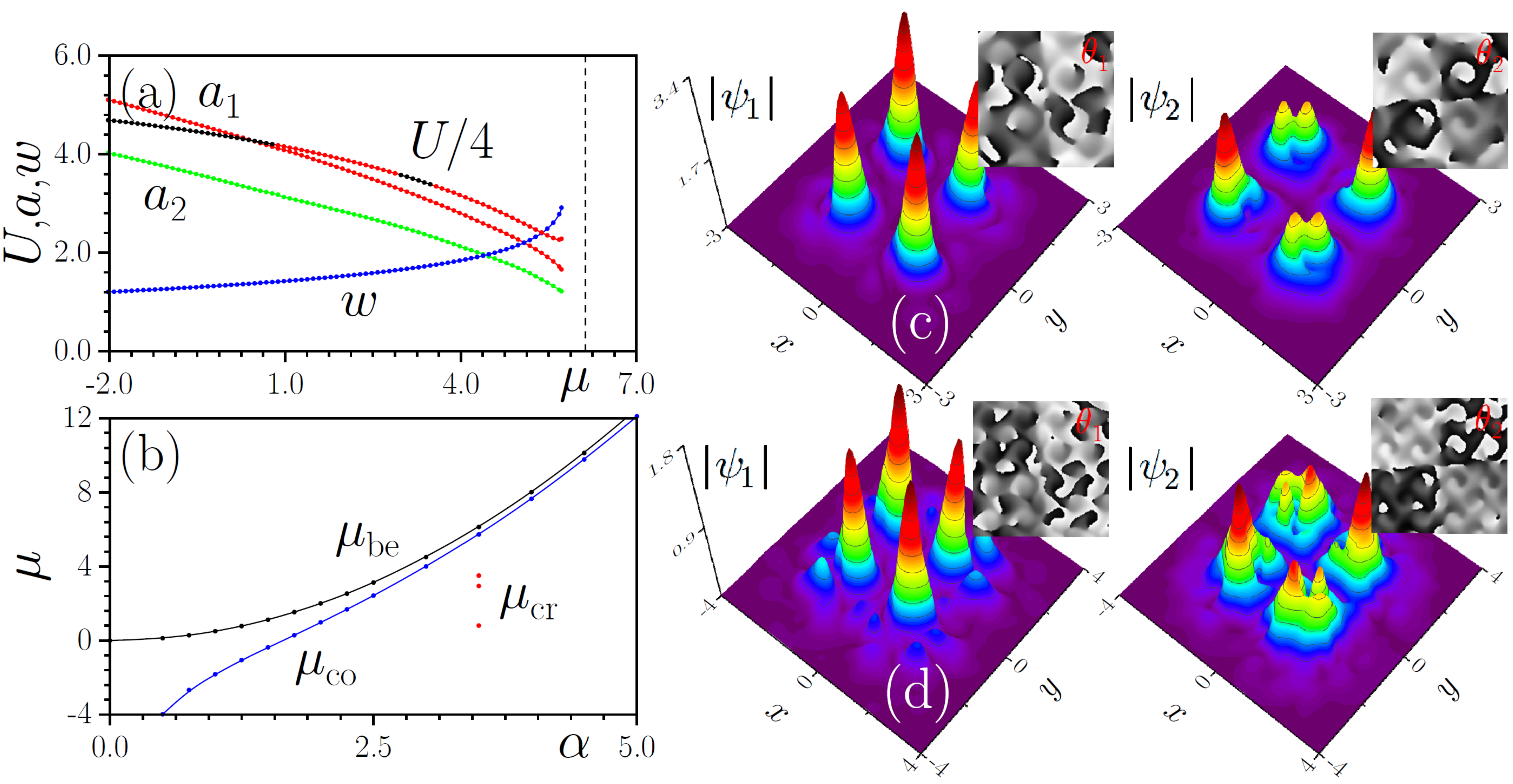}
\caption{(a) The norm $U(\mu )$, size $w(\mu )$, and
component amplitudes, $a_{1,2}(\mu)$, of quadrupole solitons versus chemical
potential $\mu $ at $\alpha =3.5$. Black and red segments in
the $U(\mu )$ dependence correspond to stable and unstable states,
respectively. (b) Existence and stability domains of the quadrupole solitons
in the $(\alpha ,\mu )$ plane. They exist at $\mu %
\leq \mu _{\mathrm{co}}<\mu _{\mathrm{be}}$. Since the shape
of the stability domain is very complex, in panel (b) we only show borders
of the stability domain at $\alpha =3.5$. Panels (c) and (d) display
examples of profiles of the absolute value and phase (insets) of quadrupoles
with $\mu =3$ (c, stable) and $\mu =5.6$ (d, unstable) for $%
\alpha =3.5$. For convenience for quadrupole solitons the SOC
profile was shifted by half of the period $\pi/\lambda$
along both $x$- and $y$-axes.}
\label{figure6}
\end{figure}

Finally, we have found quadrupole soliton complexes, composed of four
solitons with alternating signs of ${\bm\psi}$, residing at neighboring 
maxima of $\theta(x,y)$. Examples of quadrupoles and properties of their families are
presented in figure \ref{figure6}. Note that in such states diagonal pairs of
solitons have identical signs. In the course of the evolution, such pairs of
mutually attractive solitons may fuse, destroying the whole structure. For
this reason, such states are more fragile than their dipole counterparts,
and the stability region for them is narrower, with a more complex shape.
Quadrupoles also exist below the respective cutoff value of the chemical
potential, $\mu \leq \mu _{\mathrm{co}}$, see figure \ref{figure6}(a,b).

Due to the complexity of the stability domain, we were able to identify it
only for a limited set of values of the SOC strengths, $\alpha .$ For
instance, in figure \ref{figure6}(b) the quadrupoles are stable between two
upper red dots, and also below the lower one. Stable segments of the $U(\mu
) $ dependence in figure \ref{figure6}(a) are black ones. This stability was
found solely in a limited range, $3\lesssim \alpha\lesssim 4$, but not for
larger SOC strengths.

Finally, in addition to the dipoles and quadrupoles, it is possible to
construct multipole states arranged into lines, which are more stable than
the quadrupoles.

\section{Conclusions}

In this work, we have introduced a 2D system which includes the cubic
self-attraction and SOC (spin-orbit coupling) subject to the
spatially-periodic modulation, as a physically relevant dynamical model of
BEC in the binary ultracold bosonic gas. The system gives rise to
two-component (pseudo-spinor) fundamental solitons in a vast region of the
corresponding parameter space. It also essentially expands the variety of
stable self-trapped modes in the 2D setting by the creation of families of
robust dipole and quadrupole states built of two or four fundamental-like
solitons with alternating signs of the wavefunction. Stability of these solutions and properties
of their families, such as the norm, widths, and amplitudes, have been
analyzed, parameterizing the families by the chemical potential. We have
also shown that, for each species of the localized state (fundamental,
dipole, or quadrupole), additional solutions can be generated from a given
one by symmetry operations forming an Abelian eight-element group.

These results may be extended by considering axially symmetric (radial)
patterns of the spatial SOC modulation, instead of the periodic lattice
patterns. In the axisymmetric system, it may be interesting, in particular,
to consider azimuthal mobility of solitons and interactions between them. A
challenging possibility is to develop a three-dimensional generalization of
the system with the spatially periodic SOC modulation.

\textbf{Acknowledgements}

B.A.M. is supported, in part, by the Israel Science Foundation through grant
No. 1286/17, and by grant No. 2015616 from the joint program of Binational
Science Foundation (US-Israel) and National Science Foundation (US). V.V.K.
acknowledges financial support from the Portuguese Foundation for Science
and Technology (FCT) under Contract no. UIDB/00618/2020. E.Y.S. acknowledge
support by the Spanish Ministry of Science and the European Regional
Development Fund through PGC2018-101355-B-I00 (MCIU/AEI/ FEDER, UE) and the
Basque Country Government through Grant No. IT986-16. This work was
partially supported by the program 1.4 of Presidium of the Russian Academy of Sciences
"Topical problems of low temperature physics".


\section*{References}

\begin{thebibliography}{99}
\bibitem{Dressel} Dresselhaus G 1955 Spin-orbit coupling effects in zinc
blende structures Phys. Rev. \textbf{100} 580

\bibitem{Rashba} Bychkov Y A and Rashba E I 1984 Oscillatory effects and the
magneto-sussceptibility of carriers in inversion layers J. Phys. C \textbf{17%
} 6039

\bibitem{SOC1} Fabian J, Matos-Abiague A, Ertler C, Stano P and \v{Z}utic I
2007 Semiconductor spintronics Acta Physica Slovaca \textbf{57} 565

\bibitem{SOC2} Manchon A, Koo H G, Nitta J, Frolov S M and Duine R A 2015
New perspectives for Rashba spin-orbit coupling Nature Materials \textbf{14}
871

\bibitem{SOC-BEC1} Galitski V and Spielman I B 2013 Spin-orbit coupling in
quantum gases Nature \textbf{494} 49

\bibitem{SOC-BEC2} Goldman N, Juzeli\-{u}nas G, \"{O}hberg P and Spielman I
B 2014 Light-induced gauge fields for ultracold atoms Rep. Progr. Phys.
\textbf{77} 126401

\bibitem{SOC-BEC3} Zhai H 2015 Degenerate quantum gases with spin--orbit
coupling: a review Rep. Progr. Phys. \textbf{78} 026001

\bibitem{Zhang_2016} Zhang Y, Mossman M E, Busch T, Engels P and Zhang C
2016 Properties of spin-orbit-coupled Bose-Einstein condensates Front. of
Phys. \textbf{11} 118103

\bibitem{EPL} Malomed B A 2018 Creating solitons by means of spin-orbit
coupling EPL \textbf{122} 36001 (An invited Perspective mini-review;
included in Highlights of 2018)

\bibitem{1Dsol} Achilleos V, Frantzeskakis D J, Kevrekidis P J and
Pelinovsky D E 2013 Matter-wave bright solitons in spin-orbit coupled
Bose-Einstein condensates Phys. Rev. Lett. \textbf{110}, 264101

\bibitem{1Dsol2} Kartashov Y V, Konotop V V and Abdullaev F Kh 2013 Gap
solitons in a spin-orbit-coupled Bose-Einstein condensate Phys. Rev. Lett.
\textbf{111} 060402

\bibitem{1Dsol3} Xu Y, Zhang Y and Wu B 2013 Bright solitons in
spin-orbit-coupled Bose-Einstein condensates Phys. Rev. A \textbf{87} 013614

\bibitem{1Dsol4} Salasnich L and Malomed B A 2013 Localized modes in dense
repulsive and attractive Bose-Einstein condensates with spin-orbit and Rabi
couplings Phys. Rev. A \textbf{87} 063625

\bibitem{1Dsol5} Kartashov Y V, Konotop V V and Zezyulin D A 2014
Bose-Einstein condensates with localized spin-orbit coupling: Soliton
complexes and spinor dynamics Phys. Rev. A \textbf{90} 063621

\bibitem{KartKon2017} Kartashov Y V and Konotop V V 2017 Solitons in
Bose-Einstein Condensates with Helicoidal Spin-Orbit Coupling Phys. Rev.
Lett. \textbf{118} 190401

\bibitem{Kartashov_2019_2} Kartashov Y V, Konotop V V, Modugno M and Sherman
E Ya 2019 Solitons in inhomogeneous gauge potentials: Integrable and
nonintegrable dynamics Phys. Rev. Lett. \textbf{122} 064101

\bibitem{SOC2D} Wu Z, Zhang L, Sun W, Xu X-T, Wang B-Z, Ji S-J, Deng Y, Chen
S, Liu X-J and Pan J-W 2016 Realization of two-dimensional spin-orbit
coupling for Bose-Einstein condensates Science \textbf{354} 83

\bibitem{gap-sol} V. E. Lobanov, Y. V. Kartashov and V. V. Konotop 2014
Fundamental, multipole, and half-vortex gap solitons in spin-orbit coupled
Bose-Einstein condensates Phys. Rev. Lett. \textbf{112} 180403

\bibitem{Berge} L. Berg\'{e} 1998 Wave collapse in physics: principles and
applications to light and plasma waves Phys. Rep. \textbf{303} 259

\bibitem{Fibich} Fibich G \textit{The Nonlinear Schr\"{o}dinger Equation:
Singular Solutions and Optical Collapse} (Springer: Heidelberg, 2015)

\bibitem{Dias_2015} Dias J-P, Figueira M and Konotop V V 2015 Coupled
Nonlinear Schr\"{o}dinger Equations with a Gauge Potential: Existence and
Blowup Stud. Appl. Math. \textbf{136} 241

\bibitem{SpecialTopics} Malomed B A 2016 Multidimensional solitons:
Well-established results and novel findings Eur. Phys. J. Special Topics
\textbf{225} 2507

\bibitem{Kartashov2019_1} Kartashov Y V, Astrakharchik G E, Malomed B A and
Torner L 2019 Frontiers in multidimensional self-trapping of nonlinear
fields and matter Nature Reviews Physics \textbf{1} 185

\bibitem{Baizakov_2004} Baizakov B B, Malomed B A and Salerno M 2004
Multidimensional solitons in a low-dimensional periodic potential Phys. Rev.
A \textbf{70} 053613

\bibitem{Ziad} Yang J and Musslimani Z H 2003 Fundamental and vortex
solitons in a two-dimensional optical lattice Opt. Lett. \textbf{28} 2094

\bibitem{supervort} Sakaguchi H and Malomed B A 2005 Higher-order vortex
solitons, multipoles, and supervortices on a square optical lattice.
Europhys. Lett. \textbf{72} 698

\bibitem{Gaeta} Vuong L T, Grow T D, Ishaaya A, Gaeta A L, 't Hooft G W,
Eliel E R and Fibich G 2006 Collapse of optical vortices Phys. Rev. Lett.
\textbf{96} 133901

\bibitem{Malomed_2019_1} Malomed B A 2019 Vortex solitons: Old results and
new perspectives Physica D \textbf{399} 108

\bibitem{Sakaguchi_2014} Sakaguchi H, Li B and Malomed B A 2014 Creation of
two-dimensional composite solitons in spin-orbit-coupled self-attractive
Bose-Einstein condensates in free space Phys. Rev. E \textbf{89} 032920

\bibitem{Mardonov_2015} Mardonov Sh, Sherman E Ya, Muga J G, Wang H-W, Ban Y
and Chen X 2015 Collapse of spin-orbit-coupled Bose-Einstein condensates
Phys. Rev. A \textbf{91} 043604

\bibitem{Manakov_1973} Manakov S V 1973 On the theory of two-dimensional
stationary self-focusing of electromagnetic waves Zh. Eksp. Teor. Fiz.
\textbf{65} 505 [English translation: Sov. Phys. JETP \textbf{38}, 248
(1974)]

\bibitem{Chiao_1964} Chiao R Y, Garmire E and Townes C H 1964 Self-trapping
of optical beams Phys. Rev. Lett. \textbf{13} 479

\bibitem{Sakaguchi_2016} Sakaguchi H, Sherman E Ya and Malomed B A 2016
Vortex solitons in two-dimensional spin-orbit coupled Bose-Einstein
condensates: Effects of the Rashba-Dresselhaus coupling and the Zeeman
splitting Phys Rev. E \textbf{94} 032202

\bibitem{Sakaguchi_2018} Sakaguchi H and Malomed B A 2018 One- and
two-dimensional gap solitons in spin-orbit-coupled systems with Zeeman
splitting Phys. Rev. A \textbf{97} 013607

\bibitem{Li_2017} Li Y, Luo Z, Liu Y, Chen Z, Huang C, Fu S, Tan H and
Malomed B A 2017 Two-dimensional solitons and quantum droplets supported by
competing self- and cross-interactions in spin-orbit-coupled condensates New
J. Phys. \textbf{19} 113043

\bibitem{Lee_1957} Lee T D, Huang K and Yang C N 1957 Eigenvalues and
eigenfunctions of a Bose system of hard spheres and its low-temperature
properties Phys. Rev. \textbf{106} 1135

\bibitem{Petrov_2015_1} Petrov D S 2015 Quantum mechanical stabilization of
a collapsing Bose-Bose mixture Phys. Rev. Lett. \textbf{115} 155302

\bibitem{Petrov_2016_1} Petrov D S and Astrakharchik G E 2016 Ultradilute
low-dimensional liquids Phys. Rev. Lett. \textbf{117} 100401

\bibitem{Ferrier_2016}  Ferrier-Barbut I, Kadau H, Schmitt M, Wenzel M and Pfau T 2016 
Observation of quantum droplets in a strongly dipolar bose gas Phys. Rev. Lett. \textbf{116} 215301

\bibitem{Chomaz_2016} Chomaz L, Baier S, Petter D,  Mark  M J, Wachtler F, Santos L and Ferlaino F 2016 
Quantum-uctuation-driven crossover from a dilute Bose-Einstein condensate to a macrodroplet in a dipolar quantum fluid
Phys. Rev. X \textbf{6} 041039 

\bibitem{Cabrera_2018}  Cabrera C, Tanzi L, Sanz J, Naylor B, Thomas P, Cheiney P and Tarruell L 2018
Quantum liquid droplets in a mixture of Bose-Einstein condensates Science \textbf{359} 301 

\bibitem{Semengini_2018} Semeghini G, Ferioli G, Masi L, Mazzinghi C, Wolswijk L, Minardi F, Modugno M, Modugno G, Inguscio M and Fattori M 2018
Self-Bound Quantum Droplets of Atomic Mixtures in Free Space
Phys. Rev. Lett. \textbf{120} 235301

\bibitem{Zhang_2015} Zhang Y-C, Zhou Z-W, Malomed B A and Pu H 2015 Stable
solitons in three-dimensional free space without the ground state:
Self-trapped Bose-Einstein condensates with spin-orbit coupling Phys. Rev.
Lett. \textbf{115}, 253902

\bibitem{PRR} Kartashov Y V, Torner L, Modugno M, Sherman E Ya, Malomed B A
and Konotop V V 2020 Multidimensional hybrid Bose-Einstein condensates
stabi-lized by lower-dimensional spin-orbit coupling Phys. Rev. Res. \textbf{%
2} 013036

\bibitem{Sandy} Li Y, Zhang X, Zhong R, Luo Z, Liu B, Huang C, Pang W and
Malomed B A 2019 Two-dimensional composite solitons in Bose-Einstein
condensates with spatially confined spin-orbit coupling Comm. Nonlin. Sci.
Num. Sim. \textbf{73} 481

\bibitem{Chick} Garc\'ia-Ripoll J J, P\'erez-Garc\'ia V M and Vekslerchik V
2001 Construction of exact solutions by spatial translations in
inhomogeneous nonlinear Schr\"odinger equations Phys. Rev. A \textbf{64}
056602

\bibitem{Merkl} M. Merkl, A. Jacob, F. E. Zimmer, P. \"Ohberg and L. Santos
2010 Chiral Confinement in Quasirelativistic Bose-Einstein Condensates Phys.
Rev. Lett. \textbf{104} 073603

\bibitem{KartKon2020} Kartashov Y V and Konotop V V 2020 Stable Nonlinear
Modes Sustained by Gauge Fields Phys. Rev. Lett. \textbf{125}, 054101

\bibitem{Kartashov_2015} Kartashov Y V, Malomed B A, Konotop V V, Lobanov V
E and Torner L 2015 Stabilization of spatiotemporal solitons in Kerr media
by dispersive coupling Opt. Lett. \textbf{40} 1045

\bibitem{Abramowitz} Abramowitz M and Stegun I  
\textit{Handbook of Mathematical Functions: with Formulas, Graphs, and Mathematical Tables}
(Dover Books on Mathematics, 1965)

\bibitem{Knox} Knox R S and Gold A \textit{Symmetry in the Solid State}
(W. A. Benjamin, New York, 1964)

\bibitem{Kurzweil} Kurzweil H and Stellmacher B \textit{The Theory of Finite
Groups: An Introduction} (Springer, Heidelberg, 2004)

\bibitem{ZezKon} Zezyulin D A and Konotop V V 2013 Stationary modes and
integrals of motion in nonlinear lattices with a $\mathcal{PT}$-symmetric
linear part J. Phys. A: Math. Theor. \textbf{46} 415301

\bibitem{VK} Vakhitov N G and Kolokolov A A 1973 Stationary solutions of the
wave equation in a medium with nonlinearity saturation Radiophys. Quantum
Electron. \textbf{16} 783


\end{thebibliography}
\end{document}